\documentclass[prd,showkeys,superscriptaddress,nofootinbib,floatfix, 11pt]{revtex4-2}

\usepackage[colorlinks=true,
breaklinks=true,
urlcolor=magenta,
citecolor=blue]{hyperref}

\usepackage{orcidlink}
\usepackage{amssymb,amsmath,epsfig}

\usepackage{graphicx}
\usepackage{float}
\usepackage[utf8]{inputenc}
\usepackage{xcolor, soul}

\usepackage{amsmath}
\usepackage{indentfirst}
\allowdisplaybreaks[4]
\usepackage{multirow}

\newcommand{\RS}{\mathcal{R}}

\begin{document}
\title{\bf Karmarkar-Tolman Embedded Charged Anisotropic Stars in $f({\RS})$ Gravity}
\author{W.~U.~Rahman}
\email{ waheed2233@hnu.edu.cn}
\affiliation{School of Physics and Electronics, Hunan University, Changsha 410082, China}
\author{M.~Ilyas}
\email{ilyas_mia@yahoo.com}
\affiliation{Institute of Physics, Gomal University, Dera Ismail Khan, 29220, KP, Pakistan}
\author{Yi~Zhong}
\email{zhongy@hnu.edu.cn}
\affiliation{School of Physics and Electronics, Hunan University, Changsha 410082, China}
\author{De-Liang~Yao}
\email{ yaodeliang@hnu.edu.cn}
\affiliation{School of Physics and Electronics, Hunan University, Changsha 410082, China}
\affiliation{Hunan Provincial Key Laboratory of High-Energy Scale Physics and Applications, Hunan University, 410082 Changsha, China}

\date{\today}

\begin{abstract}
We investigate various anisotropic spherical distributions of charged celestial bodies within the context of $f(\mathcal{R})$ gravity, where $\mathcal{R}$ represents the Ricci scalar. The properties of specific charged compact objects are analyzed by using the Karmarkar-Tolman spacetime and three distinct gravitational models. The behavior of the structural parameters is examined via graphical methods. Energy constraints are applied to assess how well the results align with the Karmarkar-Tolman spacetime model. The physical acceptability of the stellar models is evaluated by checking the energy conditions and the equation of state parameter. Additionally, we explore the influence of anisotropy on the stability and internal structure of the models. Our findings are compared with predictions from general relativity to highlight the effects of \( f(\RS) \) gravity on charged compact stars. The obtained results are useful to enhance our understanding of how modified gravity theories affect the properties of compact astrophysical objects.

\end{abstract}
\maketitle

\newpage

\newpage

\section{Introduction}
Astrophysical compact star objects, including white dwarfs, massive neutron stars, and black holes, which originate from gravitational collapse driven by relativistic degeneracy pressure, play a crucial role in astrophysics. These celestial bodies possess exceptionally high densities because of their substantial masses, despite their relatively small volumes. Although their exact properties remain uncertain, compact stars are generally characterized as massive stars with extremely small radii; except for black holes, all such objects are classified as degenerate stars~\cite{Tolman:1939jz}.
Studying the equilibrium configurations of compact stars within the framework of general relativity (GR) is of great significance, given their high densities and massive nature. Consequently, research on compact stars, incorporating both GR and modified gravitational theories (MGTs), remains a vital and evolving field in astrophysics. So modeling these objects requires obtaining exact solutions to the Einstein field equations.
In this regard, various gravitational theories have been proposed to derive complex and precise solutions to the Einstein field equations. 

In particular, Tolman and Oppenheimer, within the context of observational data, investigated realistic models, demonstrating that the physical characteristics of these objects are the result of the interaction between internal pressure and gravitational forces, ultimately leading to equilibrium structures~\cite{Oppenheimer:1939ne}. This phenomenon, which frequently produces accurate results, is very important for studying the internal configurations of stars. Furthermore, after observing strongly magnetized spinning neutrons, Baade and Zwicky carried out a comprehensive analysis of compact stars  and found that supernovae may develop into smaller compact objects \cite{Baade:1934wuu}. In astrophysical research, the concept of a perfect fluid has been proposed as a precursor to the formation of compact stars  and the emergence of stellar bodies. These objects are typically described as ultra-dense and isotropic, exhibiting an apparent spherical symmetry. Although isotropy is often regarded as a favorable property, it is not an intrinsic characteristic of all compact stars. The concept of non-zero anisotropy in stellar structures was first introduced by Bowers and Liang \cite{Bowers:1974tgi}. Additionally, Ruderman \cite{Ruderman:1972aj} observed that the nuclear density at the core of stellar objects demonstrates anisotropic behavior. He further investigated the possibility that stars could attain extremely high densities, i.e. $10^{15}g/cm^3$, when nuclear matter exhibits anisotropic properties.

In the modified theory of gravity, an additional function \( f(\RS) \), which can include higher-order derivative terms such as \( \RS^2 \) and others, is incorporated into the Einstein-Hilbert (EH) action. Since these terms solely influence the geometric sector of the EH action, it follows that dark components are, at least theoretically, purely geometric in nature \cite{Athar:2022lxw}. MGTs have emerged as a compelling and active field of study in relativistic astrophysics in recent years. Qadir et al.~\cite{Qadir:2017lcn} emphasized the importance of incorporating advanced dynamical frameworks in research, suggesting that such approaches could offer insights into resolving dark matter (DM)-related challenges. Various MGTs, such as \( f(G) \), \( f(G,T) \), \( f(\RS) \), and \( f(G,T^2) \), have been explored, where \( G \) denotes the Gauss-Bonnet term and \( T \) represents the trace of the energy-momentum tensor. Recently, several mathematical methodologies integrating these concepts have been introduced to enhance our understanding of cosmic phenomena \cite{Nojiri:2005jg,Capozziello:2005tf,Capozziello:2011et,Odintsov:2013iba,Yousaf:2016lls,Yousaf:2021xex,Xu:2016rdf}. In these works, it is proposed that MGTs could serve as a tool for uncovering various mysteries of our universe. Theoretical studies suggest that during periods of high density, the pressure inside a star is likely to become anisotropic \cite{Ruderman:1972aj}. In recent years, numerous researchers have conducted in-depth investigations into the role of pressure anisotropy in shaping the structure of the universe \cite{Herrera:1997plx,Sharif:2014qfa,Yousaf:2017hsq,Nashed:2021gkp,Nashed:2021sji,Nashed:2022zxm,Muller:2014qja}. The structural properties of compact matter models can be significantly influenced by pressure anisotropy. To enhance our understanding of cosmic phenomena, mathematical techniques that integrate these concepts have been introduced.

It is widely recognized that in locally anisotropic systems, the ratio \({2M}/{R}\) (where \(M\) represents mass and \(R\) denotes radius) can approach unity~\cite{Karmakar:2007fn}. However, in isotropic models, the Buchdahl limit imposes a constraint, ensuring that \(\frac{2M}{R} < \frac{8}{9}\) \cite{Yazadjiev:2011ks}. Yazadjiev explored the equations of motion incorporating electric charge to analyze the hydrodynamics of relativistic systems. His findings suggested that the resulting solutions exhibited characteristics of a non-perturbative magnetar. The presence of a strong magnetic field ultimately induces anisotropy in the matter distribution. By accounting for the effects of the magnetic field in relativistic interiors, ideal fluid configurations can be utilized to construct precise analytical models. Astashenok et al.~\cite{Astashenok:2014gda}, examined the impact of strong magnetic fields on neutron star populations within the framework of modified Einstein gravity. Their study concluded that the inclusion of quadratic Gauss-Bonnet terms could support more massive neutron stars with a significant strangeness fraction. Recently, gravitational wave observations of a compact binary coalescence have identified a compact object with a mass ranging between \(2.50 M_{\odot}\) and \(2.67 M_{\odot}\)~\cite{Ilyas:2020nvt}. They investigated the upper limits of neutron stars using the causal equation of state (EoS), framework by numerically solving the corresponding Tolman-Oppenheimer-Volkoff (TOV) equations. It is well known that a neutron star will eventually collapse into a black hole if its baryon mass exceeds the stability threshold for static configurations. Our understanding of the challenging mass gap region in such systems has advanced following the work of Astashenok et al. \cite{Astashenok:2021btj}, which explored the formation of these unique structures within \( \RS^2 \) gravity. Furthermore, several other studies have examined the impact of MGTs on the potential modeling of stellar structures~\cite{Bhatti:2018iaf,Sharif:2015uqa,Sharif:2015vma,Sharif:2014ata,Sharif:2015jaa}. In a newly proposed scalar covariant theory, as well as in vector and tensor gravity models, variations in \(\omega\), \(\mu\), and \(G\) are permitted. This theory suggests that since the modified gravitational acceleration law can be derived from the equations of motion, galaxy rotation curves and cluster data can be explained without invoking DM. The concept has been successfully tested within our solar system, prompting an investigation into the linear evolution of perturbations in relation to cosmic microwave background measurements. However, scalar, vector, and tensor gravity theories face several challenges, such as the need to align with observational data, issues related to fine-tuning, and potential violations of the equivalence principle. Additionally, these theories exhibit high sensitivity to parameter variations and are inherently complex. As a result, various gravitational models distinct from GR have emerged over time, offering alternative perspectives on gravitational interactions~\cite{Capozziello:2011et,Odintsov:2013iba}. Einstein’s GR struggles to account for the late-time accelerated expansion of the universe without introducing an unknown dark energy component into the field equations~\cite{Perlmutter:2003kf,WMAP:2003ivt,Copeland:2006wr}.

In the TOV limit, the maximum mass of neutron stars is a crucial subject in astrophysics, as it determines the threshold separating neutron stars from black holes. A theoretical model predicting neutron star masses considerably exceeding observed values would contradict both empirical data and the fundamental physics of dense matter~\cite{Oppenheimer:1939ne,NANOGrav:2019jur}. The maximum possible mass of a neutron star is dictated by the TOV equation, derived from GR for a spherically symmetric object in hydrostatic equilibrium,
\begin{align}\nonumber
    \frac{dP}{dr} = - \frac{G (\epsilon + P) (m + 4\pi r^3 P)}{r (r - 2 G m)}\ 
\end{align}
where $P$ represents pressure, \( \epsilon \) is the energy density, \( m(r) \) is the enclosed mass within a radius \( r \), \( G \) is the gravitational constant.
To solve the TOV equation, an EoS describing neutron abundant matter is required. The maximum neutron star mass is influenced by the stiffness of the EoS.  There are two limits, a stiffer EoS supports higher maximum masses, and a softer EoS imposes lower masses. Also, astronomical observations suggest an upper mass limit for neutron stars. The most massive confirmed neutron stars indicate that realistic EoS models should support a mass range of approximately \( 2.0 - 2.3 M_\odot \), but not significantly beyond this.
For a neutron star to exceed ~\( 3 M_\odot \), an exceptionally stiff EoS would be necessary. However, most nuclear physics models, whether derived from microscopic approaches like chiral effective field theory or astrophysical observations such as gravitational wave detections, suggest that such an extreme EoS would be unrealistic due to fundamental constraints \cite{Romani:2022jhd,LIGOScientific:2017vwq,Fan:2023spm}.

In this work, the methodology in Refs.~\cite{Yousaf:2017lto,Rahman:2025rlz} is adopted to incorporate $f(\RS)$ modifications into a static configuration that represents an anisotropic fluid, using the Karmarkar–Tolman space-time. The Karmarkar-Tolman embedding technique offers a systematic way to construct physically realistic stellar models, especially when combined with electric charge and anisotropy, which are essential features of real astrophysical objects. Our work provides a comprehensive modeling of charged anisotropic compact stars within $f(\RS)$ gravity using three distinct gravitational models. By incorporating the Karmarkar-Tolman condition, we achieve an exact embedding of the spacetime geometry, ensuring more realistic and stable solutions. A notable advantage of our study is the detailed analysis of structural properties through graphical methods, which enhances the understanding of the physical behavior of these stars under different curvature scenarios. Additionally, the validation of our models through energy conditions and the TOV equilibrium equation further strengthens the physical reliability of the proposed configurations.

Previous studies often dealt with either anisotropic stars or charged configurations separately, or did not consider embedding techniques under $f(\RS)$ gravity~\cite{Kumar:2024ijh,Malik:2024nfl,Naz:2020ncs}. Our study fills this gap by integrating the Karmarkar-Tolman condition within the $f(\RS)$ gravity framework, leading to new and physically meaningful exact solutions. We explore and compare three different functional forms of $f(\RS)$, offering a broader perspective on how various curvature corrections affect the structure and stability of compact stars, unlike earlier works that usually focus on a single model. Through extensive graphical analysis of energy conditions, anisotropy, mass-radius relations, and charge distributions, our study offers a clear and detailed visualization of the results, reinforcing the stability and physical acceptability of the models. This approach provides more transparency and deeper insight compared to many previous purely analytical studies~\cite{Errehymy:2022ihw,Sharif:2023gbn}. Unlike many prior models that ignore the role of charge, we show that introducing charge not only affects the equilibrium and stability of compact stars but also helps in constructing more realistic models that align better with observed astrophysical objects. Thus, our work presents an important step forward in the modeling of realistic stellar structures in modified gravity, offering several theoretical and methodological improvements over previous literature.

This manuscript is organized as follows. In section~\ref{sec.fR.gravity}, general equations for irrotational spheres with $f(\RS)$ gravity corrections are presented. Section~\ref{sec.anisotropic.matter} examines specific solutions and discusses their characteristics to illustrate the approach. Section ~\ref{sec.matching} discusses the process of matching with Reissner-Nordström exterior metric. Section ~\ref{sec.Three.models.gravity} introduces three different models for $f(\RS)$ gravity. The physical aspects of $f(\RS)$ gravity models are discussed in details in Section ~\ref{sec.physical.gravity. models}. The conclusion will be presented in section~\ref{sec.sum}.

\section{Theoretical framework}
\subsection{Key field equation in $f(\RS)$ gravity\label{sec.fR.gravity}}

The \( f(\RS) \) theory of gravity is a generalization of Einstein's general theory of relativity, in which the gravitational action is modified by an arbitrary function \( f(\RS) \) of the Ricci scalar \( \RS \), instead of being simply proportional to \( \RS \). This modification allows for more flexibility in modeling the dynamics of spacetime, especially in contexts like cosmology and astrophysics, where modifications to GR might be necessary to explain phenomena such as the accelerated expansion of the universe or dark energy. The modified function in the $f(\RS)$ gravity theory can be expressed as a generalization of Einstein's GR; typically, it appears in the form of~\cite{Kasper:1993bi}.
\begin{equation}
I = \int {d{x^4}\sqrt { - g} \left[ {f(\RS)+{L_{e}}} + {L_{m}}\right]}\ ,
\end{equation}
where ${L_e}$ and ${L_m}$ represent the Lagrangians for electromagnetic and matter fields. The field equations derived from the action are more general than the standard Einstein field equations. The modified form reads 
\begin{equation}\label{eq1}
{f_\RS}{\RS_{\mu \nu }} - \frac{1}{2}{g_{\mu \nu }}f + ({g_{\mu \nu }}\Box - {\nabla _\mu }{\nabla _\nu }){f_\RS} = T_{\mu\nu}^\text{EM}+{T_{\alpha \beta }^\text{matter}}\ ,
\end{equation}
 where ${\nabla}_\mu $ is a covariant derivative; ${f_\RS}(\RS)$ is the shorthand notation of $\frac{\partial f(\RS)}{\partial \RS}$ operators; $T_{\alpha\beta}^\text{EM}$ is the electromagnetic stress-energy tensor and $T_{\mu\nu}^\text{matter}$ is the matter tensor. The explicit expressions of the electromagnetic stress-energy and matter tensor read
 \begin{align}
{T_{\mu \nu }^{\rm matter} }&= (\rho  + P){U_\mu }{U_\nu } - {P_t}{g_{\mu \nu }} + ({P_r} - {P_t}){V_\mu }{V_\nu }\ ,\\
{T_{\alpha\beta}^{\text{EM}}} &= F_{\alpha}^{\mu} F_{\beta\mu} - \frac{1}{4} g_{\alpha\beta} F^{\mu\nu} F_{\mu\nu}\ ,
\end{align}
where $U_\mu$ is the four-velocity of the fluid and $U_\mu U^\mu=1$; $V_\mu$ is the radial unit vector and $V_\mu V^\mu=-1$; $\rho$ is the energy density; the metric tensor $g_{\mu \nu}$ encodes the geometry of spacetime; $F^{\mu\nu}$ is the electromagnetic field strength tensor;  ${P_r}$ and ${P_t}$ are radial pressure components and transverse pressure components, respectively.
The components of $T_{\mu\nu}^{\text{matter}}$ and $T_{\alpha\beta}^{\text{EM}}$ are found out respectively as
\begin{align}
\text{Matter:}\quad &T_{0}^{0}=\rho\ ,\quad  T_{1}^{1}=P_r\ ,\quad  T_{2}^{2}=r^2P_t\label{3}\ ,\\
\text{EM:}\quad &T_{0}^{0}=\frac{Q^2}{8\pi r^4}\ ,\quad T_{1}^{1}=\frac{Q^2}{8\pi r^4}\ ,\quad T_{2}^{2}=\frac{r^2Q^2}{8\pi r^4}\ . \label{4}
\end{align}
Since we assume spherical symmetry, the time component of the current depends only on the radial coordinate \( r \). In the electromagnetic field tensor, the only non-vanishing components are \( F^{01} \) and \( F^{10} \), with the relation \( F^{01} = F^{10} \), representing the radial electric field.

\subsection{Anisotropic matter distribution in $f(\RS)$ gravity\label{sec.anisotropic.matter}}
In the framework of anisotropic matter distributions, the dynamics of self-gravitating systems, including stars, compact objects, and galaxies, can be examined within \( f(\RS) \) gravity to explore deviations from the conventional predictions of GR. Anisotropy is essential for understanding the internal structure of compact stars, where the radial pressure differs from the tangential pressure, a phenomenon that arises in extremely dense astrophysical systems due to intense gravitational fields, phase transitions, or electromagnetic influences.
The general form for the line element for dynamical spherically symmetric geometry is
\begin{equation}
{\rm d}{s^2} = e^{a(r)}{\rm d}{t^2} -e^{b(r)}{\rm d}{r^2}-{r^2}( {{\rm d}{\theta ^2} + {{r^2}{\sin }^2}\theta {\rm d}{\phi ^2}}) \ .
\end{equation}
For the metric above, the Karmarkar condition simplifies to a differential relation between the metric functions $a(r)$ and $b(r)$~\cite{ Mustafa:2020jln} in terms of the rank-4 Riemann tensor
\begin{align}
  \frac{R_{1414} R_{2323}}{R_{1212} R_{3434} + R_{1224} R_{1334}} = 1\ ,
\end{align}
leading to the differential condition on the metric components
\begin{align}
\frac{b'(r)}{1 - e^{b(r)}} = \frac{a'(r) \left(a'(r) + 2/r \right)}{2}\ .
\end{align}
One of the solutions for the $a(r)$ and $b(r)$ is given by
\begin{align}
   {e^a} = A{\left( {1 + B{r^2}} \right)^4} \ ,\quad {e^b} = 1 + 64{B^2}AC{r^2}{\left( {1 + B{r^2}} \right)^2}\ ,\label{eq.ab}
\end{align}
which will be used to explore the properties of compact stars under the influence of the Karmarkar and Tolman condition~\cite{Tolman:1939jz, karmarkar1948gravitational}. Here, $A$, $B$, and $C$ are the arbitrary constants to be determined by using some physical conditions.

In Karmarkar-Tolman spacetime, the metric functions \( e^{a(r)} \) and \( e^{b(r)} \) serve as relativistic analogs to gravitational potentials. 
This is a static, spherically symmetric line element commonly used in general relativity to describe stellar interiors. It forms the basis of many solutions to the Einstein field equations, including those incorporating embedding conditions such as the Karmarkar condition.
This constraint ensures the spacetime can be embedded in 5D flat space, making the solution physically consistent for modeling relativistic stars. Both functions together replace the single Newtonian potential, providing a complete description of curved spacetime.
To embed a 4D Riemannian space in a 5D flat space, the metric must satisfy the Karmarkar condition, which imposes a specific relationship between the metric functions $a(r)$ and $b(r)$. The functions $a(r)$ and $b(r)$ are metric potentials that depend only on the radial coordinate $r$, reflecting the symmetry and static nature of the configuration.
Tolman developed several exact solutions to Einstein's field equations under the assumption of this form of the metric, using different choices for $a(r)$, $b(r)$, pressure, and energy density.
The Karmarkar condition is a geometric constraint that must be satisfied for a 4-dimensional Riemannian manifold to be class one embedded in a 5-dimensional flat spacetime.

Now, solving the field equations in Eq.~\eqref{eq1}, we get
\begin{align}
\rho&=\frac{1}{8} \Bigg \{\frac{4 e^{-b(r)}}{r^2}  
\Bigg[ \Big[ 2r b'(r) + e^{b(r)} \big( r^2 \RS(r) + 2 \big) - 2 \Big] f'(\RS(r))+ r \Big[ f''(\RS(r)) \Big( \big( r b'(r) - 4 \big)
\RS'(r)\notag\\
&\hspace{1cm}- 2r \RS''(r) \Big)- 2r f^{(3)}(\RS(r)) \RS'(r)^2 \Big] \Bigg]  - 4 f(\RS) - \frac{Q(r)^2}{\pi r^4} \Bigg\} \ ,\label{eq.rho.1} \\
P_r &= \frac{1}{8} \Bigg\{ 4 e^{-b(r)} r^{-2} \Bigg[f'(\RS(r)) \Big( 2r a'(r) - e^{b(r)}\big( r^2 \RS(r) + 2 \big) + 2 \Big)\notag\\ 
&\hspace{1cm}+ r \big( r a'(r)+ 4 \big) \RS'(r) f''(\RS(r)) \Bigg]+ 4f(\RS)  + \frac{Q^2(r)}{\pi r^4}  
\Bigg\}\ ,\label{eq.Pr.1}\\
P_t&=\frac{1}{8} \Bigg\{  -\frac{2 e^{-b(r)}}{r} \Bigg[  
-2 \Big(  2 \Big( r f^{(3)}(\RS(r)) \RS'(r)^2  
+ r \RS''(r) f''(\RS(r)) + \RS'(r) f''(\RS(r)) \Big) \notag\\  
 &\hspace{1cm}+r a''(r) f'(\RS(r)) \Big) + a'(r) \Big( \big( r b'(r) - 2 \big) f'(\RS(r))
- 2r \RS'(r) f''(\RS(r)) \Big) - r a'(r)^2 f'(\RS(r))\notag \\  
&\hspace{1cm} + 2 b'(r) \Big( r \RS'(r) f''(\RS(r)) + f'(\RS(r)) \Big)  
+ 2r e^{b(r)} \RS(r) f'(\RS(r)) \Bigg]  
 + 4 f(\RS) - \frac{Q(r)^2}{\pi r^4} \Bigg\}\ ,  \label{eq.Pt.1}
\end{align}
where the Ricci scalar $\RS$ reads
\begin{equation}
\RS = \frac{{{{\rm{e}}^{ - b}}}}{{2{r^2}}}(4 - 4{{\rm{e}}^b} + {r^2}{a'^2} - 4rb' + ra'\left( {4 - rb'} \right) + 2{r^2}{a^{\prime \prime }})\ .
\end{equation}
With the definitions of $a$ and $b$ specified in Eq.~\eqref{eq.ab}, Eqs.~\eqref{eq.rho.1}, ~\eqref{eq.Pr.1} and \eqref{eq.Pt.1} can be recast to the following forms: 
\begin{align}
\rho&=\frac{1}{8} \Bigg(\frac{4}{r^2 } \left(64 A B^2 C r^2 \left(B r^2+1\right)^2+1\right)^{-1} 
\Bigg[ f'(\RS(r)) \Bigg( \left(r^2 \RS(r)+2\right) \Big( 64 A B^2 C r^2 \left(B r^2+1\right)^2+1 \Big)\notag\\
&+ {256 A B^2 C r^2 \left(3 B^2 r^4+4 B r^2+1\right)(64 A B^2 C r^2 \left(B r^2+1\right)^2+1)^{-1}}  
 - 2 \Bigg)\notag \\  
& + r \Bigg( f''(\RS(r)) \Bigg( {4 \RS'(r) \left(32 A B^2 C r^2 \left(B^2 r^4-1\right)-1\right)} ({64 A B^2 C r^2 \left(B r^2+1\right)^2+1})^{-1}) \notag\\
 &- 2 r \RS''(r) \Bigg)  
 - 2 r f^{(3)}(\RS(r)) \RS'(r)^2 \Bigg) \Bigg] - 4 f(\RS) - \frac{Q(r)^2}{\pi r^4}  
\Bigg) \ , \label{ro}\\
P_r &= \frac{1}{8} \Big (\frac{4}{r^2 }(\left(64 A B^2 C r^2 \left(B r^2+1\right)^2+1\right))^{-1}  
\Bigg[ f'(\RS(r)) \Bigg( -\left(r^2 \RS(r)+2\right) \Big( 64 A B^2 C r^2 \left(B r^2+1\right)^2 \notag\\  
&+1 \Big) + {16 B r^2}{(B r^2+1)^{-1}} + 2 \Bigg)  
+ r \Bigg( {8 B r^2}{(B r^2+1)^{-1}} + 4 \Bigg) \RS'(r) f''(\RS(r)) \Bigg] \notag \\
&+ 4 f(\RS) + \frac{Q(r)^2}{\pi r^4} \Bigg)\ ,  \label{Pr}\\
P_t& = \frac{1}{8} \Bigg(  -\frac{4}{r } (64 A B^2 C r^2 \left(B r^2+1\right)^2+1)^{-1} 
\Bigg[ r \RS(r) \left(64 A B^2 C r^2 \left(B r^2+1\right)^2+1\right) f'(\RS(r))\notag \\  
& + {128 A B^2 C r \left(3 B^2 r^4+4 B r^2+1\right) \left(r \RS'(r) f''(R(r))+f'(R(r))\right)}({64 A B^2 C r^2 \left(B r^2+1\right)^2+1})^{-1}\notag \\  
& + {8 B r}{(B r^2+1)^{-1}}  
\big( {\left(128 A B^3 C r^4) \left(B r^2+1\right)-1\right) f'(\RS(r))}{(64 A B^2 C r^2 )\left(B r^2+1\right)^2+1)^{-1}}\notag \\  
&-r \RS'(r) f''(\RS(r))\big)  
 - {32 B^2 r^3 f'(\RS(r))}{\left(B r^2+1\right)^{-2}}  
+ {8 B r \left(B r^2-1\right) f'(\RS(r))}{(B r^2+1)^{-2}}\notag\\
&- 2 \Big( r f^{(3)}(\RS(r)) \RS'(r)^2)
-2 (r \RS''(r) f''(\RS(r)) + \RS'(r) f''(\RS(r)) \Big)  
\Bigg]+ 4 f(\RS) - \frac{Q(r)^2}{\pi r^4} \Bigg) \ .\label{Po}
\end{align}
In order to obtain an explicit form of charge, the following EoS is assumed
\begin{equation}
{P_r} = \frac{1}{3}( \rho  - 4{B_g})
\end{equation}

where \( B_g \) represents the bag constant, which is the fixed energy density per unit volume of space. It signifies the inward pressure required to confine quarks within a finite spatial region. Utilizing this EoS, we obtain the explicit function of charge as
\begin{align}
Q^2 &= \pi r^2 \Bigg\{  -e^{-b(r)} \Bigg[  
3 r^2 a'(r) \RS'(r) f''(\RS(r)) + 6 r a'(r) f'(\RS(r))  
- r^2 b'(r) \RS'(r) f''(\RS(r))\notag \\  
&- 2 r b'(r) f'(\RS(r))  + 8 B_g r^2 e^{b(r)} - 4 r^2 e^{b(r)} \RS(r) f'(\RS(r))  
- 8 e^{b(r)} f'(\RS(r)) + 4 r^2 e^{b(r)} f(\RS(r))\notag \\  
& + 2 r^2 f^{(3)}(\RS(r)) \RS'(r)^2 + 2 r^2 \RS''(r) f''(\RS(r))  
+ 16 r \RS'(r) f''(\RS(r)) + 8 f'(\RS(r))  
\Bigg] \Bigg\}\ .
\end{align}

\subsection{Matching With Reissner-Nordström Exterior Metric\label{sec.matching}}
The process of matching with the Reissner-Nordström (RN) exterior metric ensures a seamless transition between an interior solution of Einstein’s field equations, representing a charged gravitating body, and the well-known RN metric, which characterizes the exterior spacetime of a spherically symmetric, charged, non-rotating mass. This matching is essential for developing physically realistic models of charged compact objects, such as charged stars or black holes. By enforcing this continuity, the process establishes constraints on the interior matter distribution, pressure profiles, and charge distribution, thereby affecting the stability and physical feasibility of models for charged compact stars or other exotic objects within GR~\cite{Castillo:1996jm}.
Here, we match the inner metric to the vacuum outer spherically symmetric metric defined by
\begin{equation}
{\rm d}{s^2} = \left( {1 - \frac{{2M}}{r}+ \frac{{Q^2}}{r^2}} \right){\rm d}{t^2} - {\left( {1 - \frac{{2M}}{r}+ \frac{{Q^2}}{r^2}} \right)^{ - 1}}d{r^2} - {r^2}\left( {{\rm d}{\theta ^2} + {\mathop{\rm \sin}\nolimits} {\theta ^2}d{\varphi ^2}} \right)\ ,
\end{equation}
which leads to 
\begin{align}
   A &= \frac{\left(-9 M R+5 Q^2+4 R^2\right)^4}{256 R^2 \left(R (R-2
   M)+Q^2\right)^3}\ , \nonumber\\ 
   B &=\frac{M R-Q^2}{R^2 \left(R (4 R-9 M)+5 Q^2\right)}\ ,\nonumber\\
   C &= \frac{R^4 \left(2 M R-Q^2\right)}{4 \left(Q^2-M R\right)^2}\ .\label{eq.ABC}
\end{align}

In what follows, we are interested in three typical compact stars, Her X-1, SAXJ 1808.4-3658, and 4U 1820-30. With the formulae specified in Eq.~\eqref{eq.ABC}, the values of the $A$, $B$ and $C$ constants can be readily calculated by using their estimated masses $M$ and radii $R$~\cite{Gangopadhyay:2013gha,Guver:2008gc}, which are compiled in Table~\ref{table:1}.
\begin{table}[h!]
\centering
\caption{ The constants $A$, $B$, and $C$ for the compact stars, Her X-1, SAXJ 1808.4-3658, and 4U 1820-30, with estimated masses $M$ and radii $R$.}\label{table:1}
\begin{tabular}{c |c c| c c c}
 \hline
Compact Stars  &$M$ & $R~(\text{km})$ & $A$ &$B$ &$C$ \\ [0.5ex]
\hline\
Her X-1  & $0.880M_{\odot}$ \cite{Gangopadhyay:2013gha} & $7.70$ \cite{Gangopadhyay:2013gha}  & $0.385712$ &$0.00175805$& $134.844$\\ [1ex]
\hline\
SAXJ1808.4-3658& $1.435M_{\odot}$ \cite{Gangopadhyay:2013gha} & $7.07$ \cite{Gangopadhyay:2013gha} & $0.214651$ &$0.00377681$ & $96.2771$\\ [1ex]
\hline
4U1820-30& $2.25M_{\odot}$ \cite{Guver:2008gc} & $10.0$ \cite{Guver:2008gc}  & $0.256916$ &$0.00161331$ & $197.484$\\ [1ex]
\hline
\end{tabular}
\end{table}

\subsection{ Three different $f(\RS)$ models\label{sec.Three.models.gravity}}

This section will explore three different models that can be used to study the properties of charged compact stars, including aspects like energy conditions, mass, and other key physical characteristics. The three models are described as follows.

\begin{itemize}
\item {\it Model 1.} First we will take the quadratic correction firstly proposed by Starobinsky~\cite{Starobinsky:1980te},
\begin{equation}\label{model1}
f(\RS) = \RS + \alpha {\RS^2}\ .
\end{equation}
Here, $\alpha$ is an arbitrary constant and $\RS$ represents the Ricci scalar. By setting \( \alpha = 0 \), we can recover GR. The function \( f \) consists of a linear term and a quadratic term, with the parameter \( \alpha \) determining the relative contribution of the quadratic component compared to the linear one. The behavior of \( f \) varies with different values of \( \alpha \). When \( \alpha \) is positive, the quadratic term dominates and grows more rapidly than the linear term for large \( \RS \). Conversely, if \( \alpha \) is negative, the quadratic term suppresses the overall growth of \( f \). 
\item {\it Model 2.}
We take the following viable model, in which the correction term to GR has a form of exponential like~\cite{Bamba:2008ut},
\begin{equation}\label{model2}
f(\RS) = \RS + \alpha \RS\left( {{e^{\left( { - \RS/\gamma } \right)}} - 1} \right)\ .
\end{equation}
Here, \( \alpha \) and \( \gamma \) are arbitrary constants. The function \( f \) combines a linear term \( \RS \) with a modified exponential term. Namely, it has two components, the first is the linear term \( \RS \), and the second involves \( \alpha \RS \) multiplied by the difference between the exponential decay \( e^{-\RS/\gamma} \) and 1. This form indicates that \( f \) depends on \( \RS \) through both linear and decaying exponential behavior, with \( \alpha \) controlling the amplitude and \( \gamma \) determining the rate of exponential decay.
\item{\it Model 3.}
Here, we apply some higher Ricci scalar terms to the quadratic gravity model, one of the cubic correction to GR~\cite{Astashenok:2014dja}, as like
\begin{equation}\label{model3}
f(\RS) = \RS + \alpha {\RS^2}\left( {1 + \gamma \RS} \right)\ .
\end{equation}
Here, \( \alpha \) and \( \gamma \) are arbitrary constants, where \( f \) depends on the variable \( \RS \). The first term, \( R \), indicates a linear dependence on \( \RS \), while the second term, \( \alpha \RS^2 (1 + \gamma \RS) \), represents a quadratic component that includes an additional factor involving \( \gamma \), making the term effectively cubic in \( \RS \). The constant \( \alpha \) controls the magnitude of the quadratic contribution, while \( \gamma \) adds further complexity by enhancing the dependence on higher powers of \( \RS \). This type of function can describe systems where both linear and non-linear effects influence the behavior of \( f \) as \( \RS \) increases.
\end{itemize}

\section{Physical Aspects of \( f(\RS) \) Gravity Models\label{sec.physical.gravity. models}}
An interesting way to calculate \( f(\RS) \) modifications characterized by two independent arbitrary constant parameters \( \alpha \) and \( \gamma \). In our work, we consider the values of these parameters are  \( \alpha \)=0.03, and  \( \gamma \)=0.5 for the \(f(\RS)\) gravity models \cite{Ilyas:2025bot}. The importance of these values was chosen to maintain the physical consistency of our charge compact star models and to comply with the experimental and observational limits. As for the small value of \( \alpha \), which is equal to 0.03, it aligns with our solar system observations. These include phenomena such as light deflection, perihelion precession and radar echo delays, all of which impose constraints on any deviations from GR in weak gravitational fields. While the \(\gamma\) parameter such as \(\gamma\)=0.5, gives the exponential correction in models \cite{Ilyas:2025bot}. The parameters are adjusted carefully to ensure stability, satisfy energy conditions, and avoid ghost instabilities across all the models. By selecting values consistent with both cosmological observations and the solar system, our models remain consistent with the well-tested predictions of general relativity, while  successfully describing the behavior of charge compact stars within the framework of modified \(f(\RS)\) gravity.

Here, we discuss some physical conditions which are necessary for the interior solution. In the following, we present the anisotropic behavior and stability
conditions.

\subsection{Energy Density and Pressure Evolutions}

\begin{figure}[ht] \centering
\epsfig{file=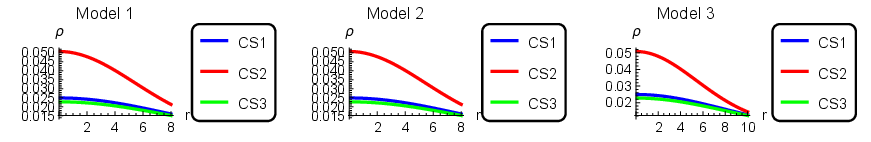,width=1\linewidth}
\caption{Plot of the density evolution of the strange star candidates, Her X-1,
SAX J 1808.4-3658 and 4U 1820-30, under three different Models.}\label{roc}
\end{figure}

\begin{figure}[ht] \centering
\epsfig{file=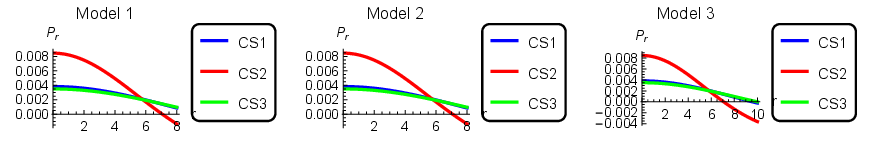,width=1\linewidth}
\caption{Graph depicting the evolution of radial pressure for the strange star candidates, Her X-1, SAX J1808.4-3658 and 4U 1820-30, under three distinct models.}\label{prc}
\end{figure}

\begin{figure}[ht] \centering
\epsfig{file=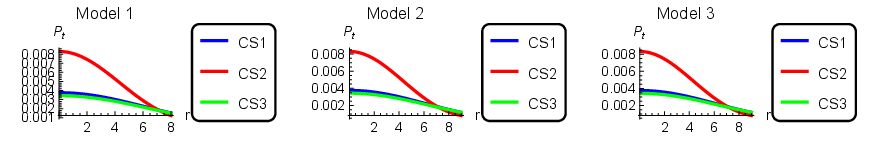,width=1\linewidth}
\caption{Graph depicting the evolution of transverse pressure for the strange star candidates, Her X-1, SAX J 1808.4-3658 and 4U 1820-30, under three distinct models.}\label{ptc}
\end{figure}

The density plot for the strange star candidates, Her X-1, SAX J 1808.4–3658 (SS1) and 4U 1820-30, as shown in Fig.~\ref{roc}, illustrates that as \( r \to 0 \), the density \( \rho \) reaches its maximum value. In simple terms, \( \rho \) behaves as a decreasing function of \( r \), meaning that as \( r \) increases, \( \rho \) decreases. This pattern signifies the high compactness of the stellar core, confirming that our \( f(\RS) \) models under investigation remain valid in the outer regions of the core.

\begin{table}[h!]
\centering
\caption{Bag Different constant values.}
\label{table:2}
\setlength{\tabcolsep}{5mm}{\begin{tabular}{l| c| c| c}
 \hline
  & \multicolumn{3}{c}{$10^3\times B_g$}   \\ 
  \cline{2-4}
Models  & Her X-1 & SAXJ1808.4-3658 & 4U1820-30 \\ 
\hline
Model 1 & 4.27539 & 5.94515 & 2.88740\\ 
\hline
Model 2 & 4.23099 & 5.83972 & 2.86436\\ 
\hline
Model 3 & 4.28107 & 5.95875 & 2.89035\\ 
\hline
\end{tabular}}
\end{table}

As it is well known, the radial pressure vanishes at the surface, i.e., \( P_r(R) = 0 \). Using the radial pressure condition, we determine the value of \( B_g \) for the three relativistic spheres, as presented in Table~\ref{table:2}. With Eqs.~\eqref{ro}-\eqref{Po} and the involved constants $A$, $B$ and $C$ given in Table~\ref{table:1}, we obtain $\rho$, $P_r$, and $P_t$. Figs.~\ref{prc} and~\ref{ptc} illustrate the variation in anisotropic radial and transverse pressures, denoted as \( P_r \) and \( P_t \), respectively.

\begin{figure} \centering
\epsfig{file=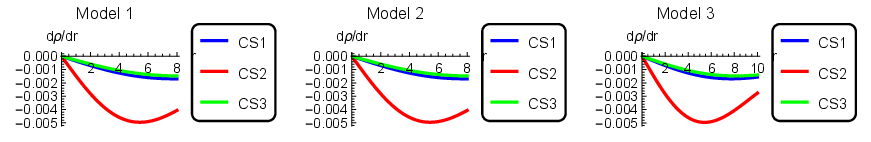,width=1\linewidth}
\caption{Plot of \( \frac{d\rho}{dr} \) as a function of increasing \( r \) for the strange star candidates Her X-1, SAX J1808.4-3658, and 4U 1820-30, considering three different models.}\label{dro}
\end{figure}

\begin{figure} \centering
\epsfig{file=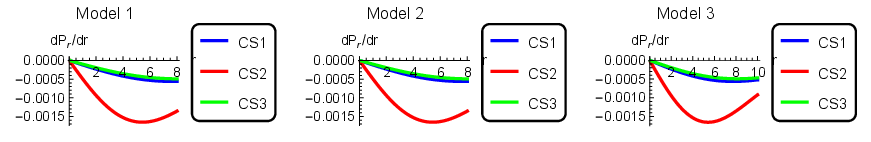,width=1\linewidth}
\caption{A graph of \( \frac{dP_r}{dr} \) as a function of increasing \( r \) for the strange star candidates, Her X-1, SAX J 1808.4-3658 and 4U 1820-30, based on three different models.}\label{dpr}
\end{figure}

\begin{figure} \centering
\epsfig{file=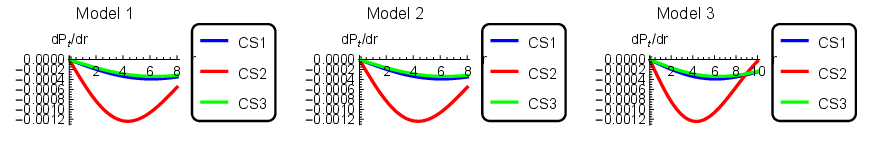,width=1\linewidth}
\caption{The graph of \( dP_t/dr \) versus \( r \) is presented for the strange star candidates, Her X-1, SAX J 1808.4-3658 and 4U 1820-30, illustrating the behavior of this quantity across three distinct models.}\label{dpt}
\end{figure}

The variation of radial derivative of density, radial and transverse pressure are shown in Figs.~\ref{dro}, \ref{dpr} and~\ref{dpt}, respectively.
We see that $\frac{d\rho}{{dr}}<0$, $\frac{dP_r}{{dr}}<0$ and $\frac{dP_t}{{dr}}<0$ for all three models and strange stars. For $r=0$, we get
\begin{align}
    \frac{d\rho}{{dr}}=0\ ,\quad \frac{dP_r}{{dr}}=0\ .
\end{align}
 As expected, we have the maximum density for the minimum value of $r$, (star core density $\rho(0)=\rho_c$) because these are the decreasing function.

\subsection{Energy conditions}
The general expression for the energy conditions can be derived from the Raychaudhuri equation for expansion~\cite{hawking2023large}. From these conditions, it can be inferred that gravity is inherently attractive, with a positive energy density that cannot exceed the speed of light. The energy conditions are defined as follows.
\begin{itemize}
    \item {Null Energy Condition (NEC):}
    \begin{align}
        P_r+\rho  \ge 0, \quad P_t+\rho  \ge 0\ .\label{eq.NEC}
    \end{align}
    \item {Weak Energy Condition (WEC):}
    \begin{align}
    \rho \ge 0, \quad \rho + P_r \ge 0, \quad \rho + P_t \ge0 \ .\label{eq.WEC}
    \end{align}
    \item {Strong Energy Condition (SEC):}
    \begin{align}
    \rho + P_r \ge 0, \quad \rho + P_t \ge 0, \quad \rho + P_r + 2P_t \ge 0\ .\label{eq.SEC}
    \end{align}
    \item {Dominant Energy Condition (DEC):}
    \begin{align}
    \rho \ge |P_r|, \quad \rho \ge |P_t|\ .\label{eq.DEC}
    \end{align}
\end{itemize}

\begin{figure}\centering
\centering
\epsfig{file=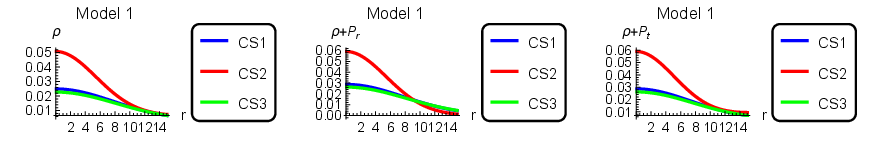,width=1\linewidth}
\epsfig{file=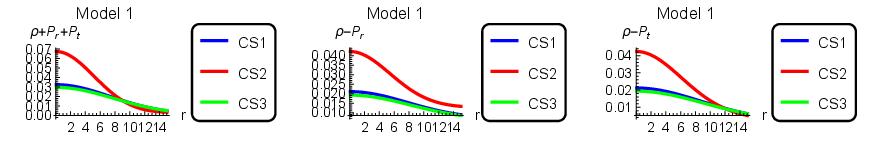,width=1\linewidth}
\caption{Different Energy conditions for strange stars, Model 1.}\label{energy1}
\end{figure}

\begin{figure}\centering
\centering
\epsfig{file=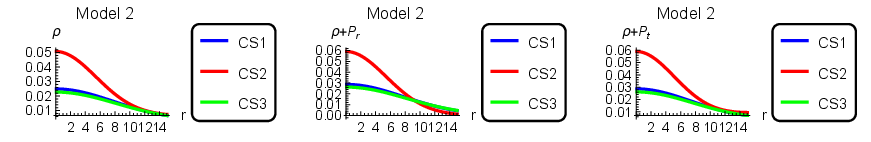,width=1\linewidth}
\epsfig{file=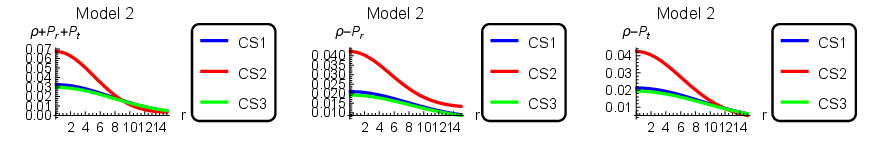,width=1\linewidth}
\caption{Different Energy conditions for strange stars, Model 2.}\label{energy2}
\end{figure}

\begin{figure}\centering
\centering
\epsfig{file=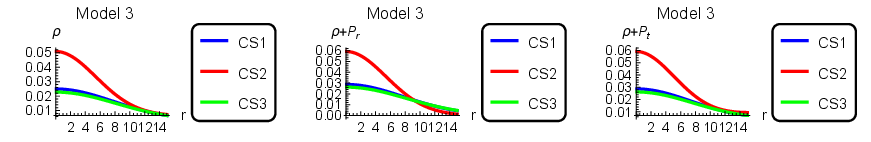,width=1\linewidth}
\epsfig{file=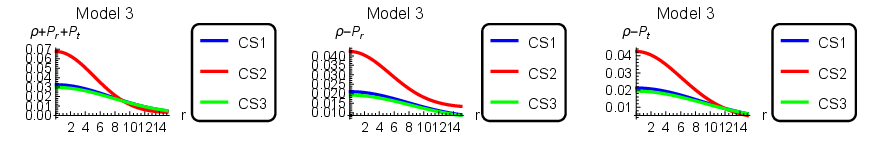,width=1\linewidth}
\caption{Different Energy conditions for strange stars, Model 3.}\label{energy3}
\end{figure}

In our study, the NEC~\eqref{eq.NEC} and SEC~\eqref{eq.SEC} have been thoroughly examined across all three considered $f(\RS)$ models, following the framework discussed in Ref.~\cite{Rahman:2022vxe}. In fact, the evolution of all the above energy conditions for compact stars is thoroughly satisfied in our viable models, as illustrated graphically for all three strange star candidates in Figs.~\ref{energy1}, \ref{energy2}, and \ref{energy3}. This confirms that the matter distribution remains physically acceptable and avoids exotic behavior such as negative energy densities.

\subsection{TOV Equation}
The modeling of stellar structure in scenarios where general relativistic effects play a significant role, such as in neutron stars, relies on solving Einstein’s equation for a spherically symmetric star in static equilibrium. This framework describes the star in terms of energy density, pressure, and a quantity analogous to the gravitational potential. The resulting equation can be seen as a natural extension of the principles governing stars in hydrostatic equilibrium~\cite{smith2012tolman}.

\begin{figure}[ht] \centering
\epsfig{file=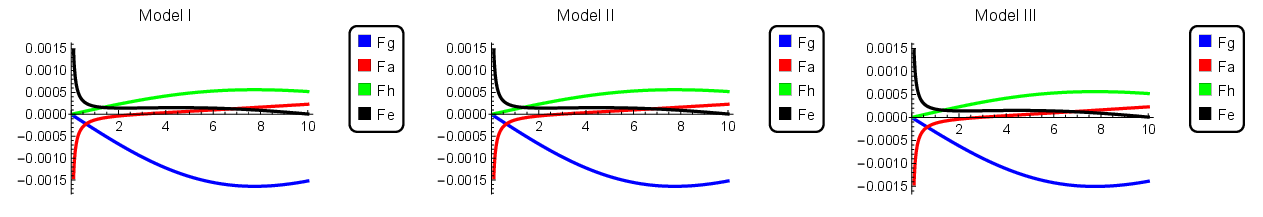,width=1\linewidth}
\caption{A graph showing the gravitational force ($F_g$), hydrostatic force ($F_h$), anisotropic force ($F_a$), and electrostatic force ($F_e$) as functions of the radial coordinate $r$ (in km).}\label{eqb}
\end{figure}

The TOV equation can be expressed in a generalized form as
\begin{equation}
\frac{dP_r}{dr} + \frac{{a'}(\rho + P_r)}{2} + \frac{2(P_r - P_t)}{r} +\frac{Q}{4 \pi r^4} \frac{dQ}{dr}=0\ .
\end{equation}
Furthermore, it can be written in form of gravitational, hydrostatic, anisotropic and electromagnetic forces (denoted by $F_g$, $F_h$, $F_a$ and $F_e$ in order)
\begin{equation}
F_g + F_h + F_a+F_e= 0,
\end{equation}
which yields
\begin{align}
    F_g=-\frac{4 B r ({P_r}+\rho )}{1+ B r^2}\ ,\quad  F_h=-\frac{{d{P_r}}}{{dr}}\ ,\quad   F_a= \frac{{2({P_r} - {P_t})}}{r}\ ,\quad F_e=\frac{Q}{4 \pi r^4} \frac{dQ}{dr}\ .
\end{align}
Using these definitions, we generate plots for three strange compact stars, as illustrated in Fig.~\ref{eqb}.

Fig.~\ref{eqb} shows the variation of the four forces with respect to the radial coordinate $r$~(km). More specifically, these plots illustrate the variation of different force components contributing to the hydrostatic equilibrium of a charge compact stellar object under three different gravitational models. As can be seen form Fig.~\ref{eqb}, the gravitational force acts inward throughout the star. It is the dominant attractive force and decreases in magnitude with increasing radius. The anisotropic force is directed inward, indicating negative anisotropy (radial pressure greater than tangential). It slightly modifies the equilibrium but remains smaller in magnitude than gravity. The hydrostatic force is positive and outward-directed. It represents the pressure gradient that resists gravitational collapse. It gradually decreases toward the surface. Finally, the electric force is strong near the center due to high charge density, then quickly diminishes. It provides an outward repulsive effect and contributes to the stability of the star. Our results confirm that the combined effects of hydrostatic, electric, and anisotropic forces balance the gravitational pull, maintaining equilibrium throughout the stellar interior. This ensure the configuration is stable and physically acceptable under the specified three different models.

\subsection{Stability Analysis}
In this section, we examine the stability of compact star models within the framework of $f(\RS)$ theory. To assess the stability of our model, we compute the radial velocity ($v_{sr}$) and transverse velocity ($v_{st}$), which are defined via
\begin{align}
    \frac{{d{P_r}}}{{d\rho }} = v_{sr}^2\ ,\quad \frac{{d{P_t}}}{{d\rho }} = v_{st}^2\ .
\end{align}
For stability, both the radial and transverse speeds must satisfy the condition, $0 \le v_{sr}^2 \le 1$, i.e., the value of  $v_{sr}^2$ is nearly equal to $\frac{1}{3}$ and $0 \le v_{st}^2 \le 1$.

\begin{figure} \centering
\epsfig{file=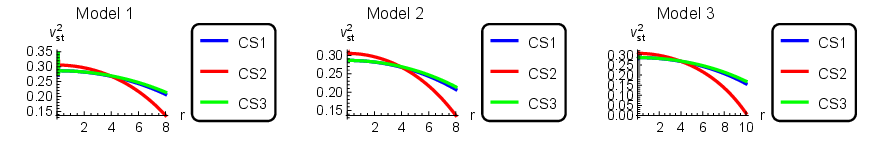,width=1\linewidth}
\caption{Variations of $v_{st}^2$ with respect radius r (km) of the strange star}\label{vst}
\end{figure}

\begin{figure}[ht] \centering
\epsfig{file=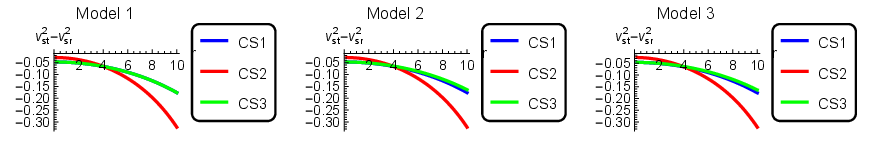,width=1\linewidth}
\caption{Variations of $v_{st}^2 - v_{sr}^2$ with respect radius $r$ (km) of the strange star}\label{vstmvsr}
\end{figure}

As shown in Fig.~\ref{vst}, the evolution of the radial and transverse sound speeds for all three types of strange star candidates remains within the stability limits, as previously discussed. The speed of sound \( v_s \) in dense matter must remain below the speed of light \( c \). Some extreme EoS models require \( v_s \approx c \), which is physically implausible. Observational data indicate a distinct mass gap between the heaviest neutron stars (\( 2.3 M_\odot \)) and the lightest black holes (\( 5 M_\odot \)). When a neutron star exceeds its maximum mass, such as through mass accretion in a binary system, it undergoes rapid gravitational collapse within milliseconds, forming a black hole. This phenomenon explains the observed absence of compact objects with masses between \( 2.5 - 5 M_\odot \). Core-collapse supernovae typically yield either neutron stars (\(\lesssim 2.5 M_\odot\)) or black holes (\(\gtrsim 5 M_\odot\)), with little space for intermediate objects. In a similar way, it can be observed from Fig.~\ref{vstmvsr} that $0< |v_{st}^2-v_{sr}^2|<1$, so the stability is attained for compact stars in our considered $f(\RS)$ models.

\subsection{EoS parameter}
The EoS parameter \( w \) is essential for understanding the thermodynamic properties of cosmic fluids and their impact on the universe large-scale structure and its evolution. In extended gravitational models beyond GR, such as \( f(\RS) \) gravity, scalar-tensor theories, and other modified gravity frameworks, the EoS parameter can vary over time, providing a dynamic view of cosmic acceleration. These models often introduce additional scalar fields that influence \( w \), resulting in its evolution and potential transitions between different values throughout cosmic history. The EoS parameter is an essential tool for describing both the universe expansion and the properties of cosmic fluids. Whether constant or time-dependent, its behavior offers crucial understanding of dark energy, the early universe, and the long-term evolution of the cosmos. By linking theoretical models with observational data, it plays a central role in modern cosmology and astrophysics.

\begin{figure} \centering
\epsfig{file=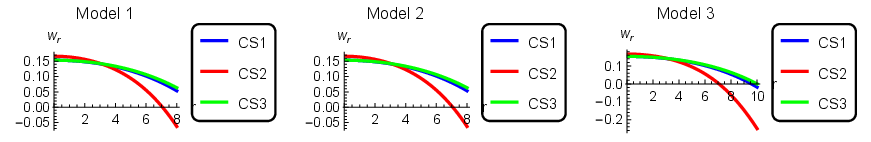,width=1\linewidth}
\caption{Variations of the radial EoS parameter with respect to the radial.}\label{wr}
\end{figure}
\begin{figure} \centering
\epsfig{file=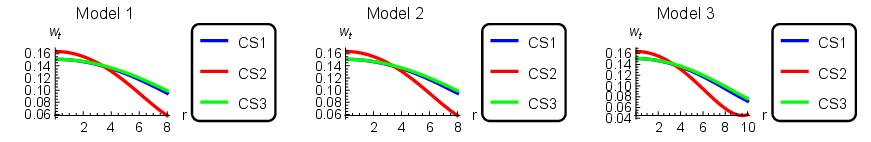,width=1\linewidth}
\caption{ Changes in the transverse equation of state (EoS) parameter in relation to the radial direction.}\label{wt}
\end{figure}

For the anisotropic case, the EoS can be expressed as
\begin{align}
    P_r=w_r \rho\ ,\quad P_t=w_t \rho\ ,
\end{align}
where $w_{r,t}$ are relevant radial and transverse EoS parameters, respectively. The limits are such that \( 0 < w_r < 1 \) and \( 0 < w_t < 1 \). The graphical representation of the behavior of \( w_r \) and \( w_t \) is shown Figs.~\ref{wr} and~\ref{wt}, where $0<w_r<1$ and $0<w_t<1$ are visible, indicating that the matter content is typical real matter.

\subsection{The Measurement of Anisotropy}
Anisotropy in compact stars, like neutron stars and quark stars, refers to the differences in pressure that occur along various directions within the star. This deviation from isotropic pressure distribution is significant in understanding the internal structure, stability, and EoS of these dense objects. Various physical mechanisms, such as strong magnetic fields, phase transitions, and anisotropic stress from exotic matter, can induce anisotropy in compact stars. The measurement of anisotropy in compact stars is typically inferred through astrophysical observations and theoretical modeling. Additionally, theoretical models employing GR, the TOV equation, and modified gravity frameworks help estimate the impact of anisotropy on mass-radius relations and maximum mass constraints. 

\begin{figure} \centering
\epsfig{file=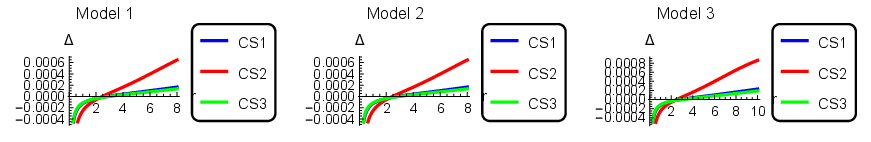,width=1\linewidth}
\caption{Variations of anisotropic measure $\Delta$ with respect to the radial.}\label{ani}
\end{figure}

The anisotropy is defined as
\begin{equation}
\Delta  = \frac{2}{r}({P_t} - {P_r})\ .
\end{equation}
Plots of the anisotropy are displayed in Fig.~\ref{ani}. We observe that $\Delta > 0$, such as $P_t > P_r$, indicating that the anisotropy is oriented outward. 

\subsection{Electric Field and Charge}  

\begin{figure}  
\centering  
\epsfig{file=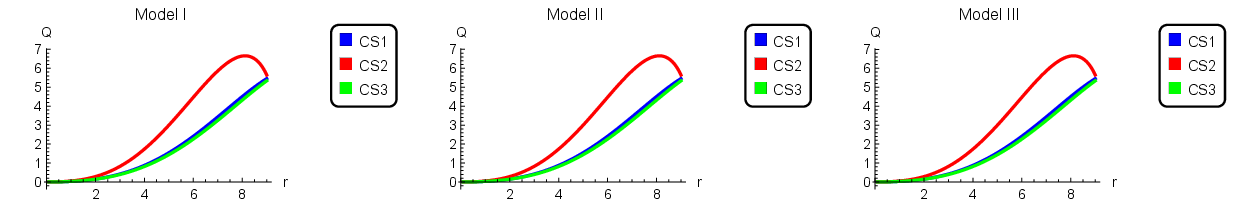,width=1\linewidth}  
\caption{Radial distribution of the charge function \( q(r) \), derived from Gauss’s law in curved spacetime, illustrating the accumulation of charge within the stellar interior.}  
\label{charg}  
\end{figure}  

\begin{figure}  
\centering  
\epsfig{file=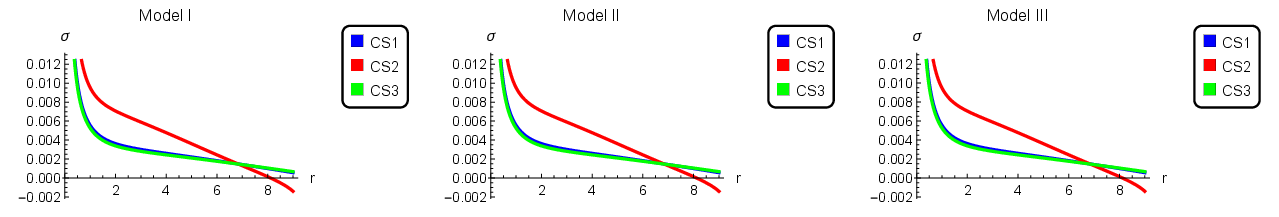,width=1\linewidth}  
\caption{Charge density \( \sigma(r) \) as a function of the radial coordinate \( r \), illustrating the spatial distribution of charge in the star.}  
\label{chargedensity}  
\end{figure}

The equilibrium configuration of a charged compact star is shaped by the interplay between gravitational attraction and electrostatic repulsion. The charge distribution \( q(r) \), determined using Gauss’s law in curved spacetime, is depicted in Fig.~\ref{charg}. Additionally, understanding the behavior of \( \sigma(r) \) is essential for determining whether charge is concentrated near the core or extends towards the outer layers of the star. The charge density \( \sigma(r) \), which describes the spatial distribution of charge within the stellar interior, is illustrated in Fig.~\ref{chargedensity}. The charge density \( \sigma(r) \), shown in Fig.~\ref{chargedensity}, is obtained from the derivative of \( q(r) \) and reveals the manner in which charge is localized within the star. These profiles provide valuable insights into how charge accumulates and varies across different regions of the star.

\begin{figure}  
\centering  
\epsfig{file=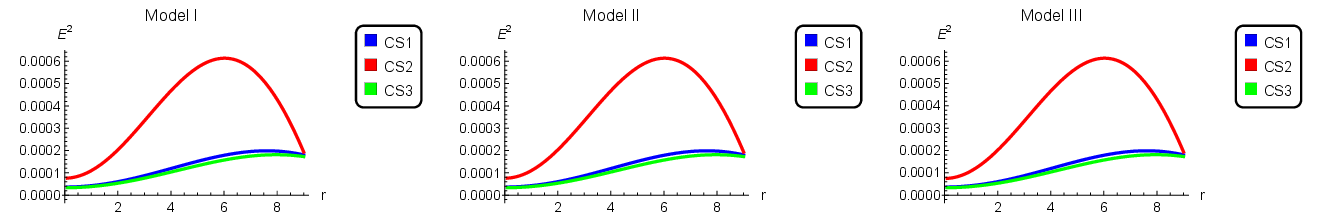,width=1\linewidth}  
\caption{Squared electric field \( E^2(r) \) as a function of radial distance, demonstrating the role of charge in maintaining the stability of the compact star.}  
\label{electric}  
\end{figure}

The presence of charge introduces an electric field that significantly alters the star structure. A strong electric field generates an outward force that counteracts gravitational collapse, thereby contributing to the equilibrium of the system. The electric field \( E^2(r) \), derived from the charge function, plays a crucial role in modifying the internal pressure gradients and influencing the overall stability of the configuration. Fig.~\ref{electric} presents the variation of the squared electric field \( E^2(r) \) as a function of radial distance. The behavior of \( E^2(r) \) provides crucial information on how charge contributes to stabilizing the stellar configuration against gravitational collapse. In Fig.~\ref{electric}, the radial dependence and peak values of \( E^2(r) \) highlight its impact on the star structure.

\section{Summary\label{sec.sum}}
\vspace{0.5cm}

In this work, we explore stable stellar models with an anisotropic matter distribution within the framework of \( f(\RS) \) gravity, considering a static and spherically symmetric configuration. The stellar metric system is formulated using the Krori-Barua solution, where junction conditions on a hypersurface determine the unknown constants. The study utilized observational data from three different star models, incorporating parameters such as mass and radius. The primary goal is to examine the impact of additional curvature terms in the \( f(\RS) \) theory on the stability of charged compact stars. To perform this analysis, three specific \( f(\RS) \) models were considered,  which explored the relationship between mass and radius in neutron stars with quadratic and exponential Ricci scalar corrections. Our findings also indicate that the incorporation of an EoS leads to a significant increase in the maximum mass of neutron stars, supporting the plausibility of \( f(\RS) \) gravity.\\  

The radial coordinate \( r \) influences physical properties such as energy density and anisotropic stresses. We show that the energy density decreases as the star radius increases, suggesting a maximally compact configuration at the core. A similar trend is observed in the behavior of both tangential and radial pressures. This observation suggests the formation of dense celestial structures. The first derivative of the pressure components transitions from negative to positive, indicating the presence of a local minimum. This highlights the compact nature of the object, a similar pattern observed for \( \frac{dP_t}{dr} \). The locally anisotropic sphere is structured to satisfy all energy conditions, at a specific radius \( r \). In the TOV equation, the forces like \(F_g\), \( F_h \), \( F_a \) and \(F_e \) suggest that all charged compact stars will eventually reach a state of hydrodynamic equilibrium. The relation between \( \omega_{r,t} \)  and $r$ within a specified range, the compact internal core of stars in $f(\RS)$ gravity is demonstrated by the fact that both \( \omega_{r} \) and \( \omega_{t} \) remain between 0 and 1.\\

  We calculate the radial and transverse components of sound speed. The transverse sound speed squared is greater than the radial sound speed squared. By calculating the difference between these two speeds, we evaluate the system stability and ensure that \( 0 < |v_{st}^2 - v_{sr}^2| < 1 \), which reduces redundancy and establishes meaningful connections between the components of our models. The anisotropy parameter \( \Delta = P_t - P_r \) stays positive across the configuration, contributing to the system resistance against gravitational collapse. The electric field \( E^2(r) \), derived from the charge function, plays a crucial role in modifying the internal pressure gradients and influencing the overall stability of the configuration. The upper mass limit is largely governed by the EoS of dense nuclear matter and the impact of general relativistic gravitational effects. It is important to note that the limitation of this study arises from the specific conditions and assumptions imposed by the selected $f(\RS)$ gravity models and Karmarkar-Tolman spacetime.

\acknowledgments
This work is supported by the National Nature Science Foundations of China (NSFC) under Contract No.~12275076, No.~12335002 and No. 12305061; by the Science Fund for Distinguished Young Scholars of Hunan Province under Grant No.~2024JJ2007; by the Fundamental Research Funds for the Central Universities under Contract No.~531118010379.

\bibliography{mybib1}
\end{document}